# Accuracy, Repeatability, and Reproducibility of a Radiographic Technique to Assess Spinal Cord Stimulation Lead Position: A Validation Study

[1]Andrew Thoreson M.S., [1,2]Katrina Fernandez B.E.E., [1]Cesar Lopez M.S.,
[1]Margaux Linde B.S., [3]Mark A. Bendel M.D., [1]Peter Grahn Ph. D, [1,2,4]Kristin D. Zhao Ph. D.

1. Assistive and Restorative Technology Laboratory, Rehabilitation Medicine Research Center, Department of Physical Medicine and Rehabilitation, Mayo Clinic, Rochester, MN, USA
2. Mayo Clinic Graduate School of Biomedical Sciences, Mayo Clinic, Rochester, MN, USA
3. Department of Anesthesiology and Perioperative Medicine, Mayo Clinic, Rochester, MN, USA
4. Department of Physiology and Biomedical Engineering, Mayo Clinic, Rochester, MN, USA

Corresponding Author:
Kristin D. Zhao
kristin.zhao@mayo.edu

This is a preprint of the article that has not gone through full peer review.

**Abstract**
Spinal cord stimulation with implantable leads is a valuable therapy used to treat a variety of chronic pain conditions. However, lead migration is a common complication causing loss of efficacy. Previous reports have characterized lead migration using radiographs, but methods are not consistent and lack rigorous validation. The purpose of this study was to develop a technique to perform radiographic measurements of the position of epidural spinal cord leads within the lumbosacral spinal canal and establish its accuracy, repeatability, and reproducibility. Computed tomography scans were acquired from three clinical trial participants implanted percutaneously with two eight-contact cylindrical leads; from these, electrode positions were established using three-dimensional measurements, and digitally reconstructed radiographs were created. Two operators applied a digitization and measurement protocol for each lead. Bland-Altman plots were created to determine smallest detectable change, and a gage repeatability and reproducibility analysis was performed. Smallest detectable change was found to be less than the distance between adjacent electrodes and variability introduced by repeatability and reproducibility was less than 10% of the total study variability. We conclude that the method developed to measure lead electrode position has sufficient accuracy and acceptable repeatability and reproducibility.

## Introduction

Spinal cord stimulation delivered with implantable leads has been a valuable therapy to assess and treat patients with chronic pain [1, 2]. Despite widespread use, complications have been reported, including lead migration. Incidence of lead migration causing complications with stimulation therapy have been reported to range from 2.1% to 27% [2] and can occur due to insufficient anchoring, unstable entry-point orientation, changes in lead tensile loading or repeated patient motions [3-6]. Solutions to restore efficacy range from conservative approaches such simple reprogramming to more complicated revision surgery all depending on the severity of the migration [4, 7].

A number of studies have described quantitative assessment of lead tip position and migration from radiographs using different measurement strategies [8-14]. Methodological details are often sparse, but key features of techniques often include manual measurements made with radiology software [8, 11], definition of consistent vertebral references [10, 13, 15], and magnification correction [10, 12-14]. Hofstoetter et al. describe a technique to radiographically assess rostral-caudal position of leads using additional MRI data to correct for out-of-plane tilt of the spinal canal [16]. Assessments of accuracy of any of these techniques are lacking in the literature.

Establishing a gold standard against which these techniques can be compared is challenging. Previous methods described for assessing lead position require human operator interaction to select key points on leads and landmarks. However, repeatability and reproducibility studies of these methods have not been reported. A validated, accurate technique for quantifying lead migration is essential for evaluating new mechanisms for securing leads in place [10] and in developing a standard characterization for lead migration to determine its true incidence rate.

The purpose of this study is to assess the accuracy, repeatability, and reproducibility of a technique to quantitatively assess neuromodulation lead position relative to spine anatomy using single plane radiographic imaging. Our hypothesis was that measurement errors from gold standard values would fall below a clinically meaningful threshold and that measurements would be highly repeatable and reproducible. A reliable, validated approach to quantitatively assess lead position will be important to understand the true incidence of lead migration.

## Methods

*Study Data*: Three participants with spinal cord injury were recruited to a clinical trial investigating the effect of neuromodulation on lower extremity outcomes following approval from the Food and Drug Administration and Mayo Clinic Institutional Review Board (Clinicaltrials.gov identifiers: NCT05095454). Participants were surgically implanted with two temporary percutaneous neuromodulation leads (Model 3086, Abbott Neuromodulation, Austin, TX) with target placement in the T11-L1 region of the lumbar enlargement of the spinal cord with leads biased from the mid-line (one to the left and one to the right) by approximately 5 mm. Final placement was guided by intraoperative anterior-posterior and medial-lateral radiograph projections and electromyography assessment of the lower extremity muscles[17]. Computed Tomography (CT) scans were obtained two days after surgery to assess lead placement and determine if significant lead migration had occurred. CT scan parameters are detailed in Table 1.

*Gold Standard Measurements*: Positions of the two leads at the post-operative timepoint were characterized from the CT data by digitizing key vertebral landmarks and lead electrodes using medical imaging analysis software (ANALYZE, Mayo Clinic, Rochester, MN). A 3D reconstruction of CT data was performed. Landmark and electrode centroid position were defined in the imaging coordinate system. Vertebral landmarks were selected from the axial slices of the CT reconstruction. For T12, L1 and L2 vertebrae, three points were digitized on each superior endplate – one on the left lateral edge, one on the right lateral edge, and the third on the anterior edge (Figure 1a). Electrode centroids were identified by bisecting their radiopaque outline with all three sectioning planes and recording intersection coordinates (Figure 1b).

*Digitally Reconstructed Radiographs*: Anterior-posterior plane digitally reconstructed radiographs (DRRs) were created from CT using MATLAB (version 2025, The Mathworks, Natick, MA). CT volume data was imported, Hounsfield unit (HU) values for voxels were summed along the projection dimension, the array was rescaled to the original range of -1024 to 3071 HU and then a grayscale image was exported in TIFF format. To attenuate the metal artifact surrounding electrodes, all values greater than 3000 HU (a value observed to be higher than typical artifact values) in the CT volume were multiplied by a factor of three, determined empirically to sufficiently reduce the noise, before planar compression.

*Radiograph Lead Measurements*: A standard protocol was developed to characterize lead position from anterior-posterior intraoperative radiographs of patients lying prone. Digitization was performed within the "Measure" module of ANALYZE with the 1-D "Sample Points" tool applied to digitize user-identified features. Beginning with the rostral-most lead electrode, Sample Points were marked and coordinates logged at the rostral and caudal boundary of the electrode near the centerline, adjusting image magnification as necessary to make an accurate selection (Figure 2). The process was to digitize all eight electrodes. Digitization was completed by placing Sample Points at the four outermost visible corners of anterior vertebral body T12, L1, and L2 projections. Data was processed into distance measurements using a MATLAB algorithm. The center of line defining each superior endplate was calculated; the vertebra with its superior endplate center closest to electrode point field centroid was defined as rostral-causal reference plane [10, 13, 15]. A perpendicular medial-lateral reference plane also crossed this center point. Centroids of points defining each electrode were calculated. Dimensions were then scaled so that the distance between the two nearest electrode centroids was equal to nominal electrode center-to-center spacing of 7 mm [12-14]. A position vector between each electrode centroid and reference origin was resolved into rostral-caudal (Z, rostral positive) and medial-lateral (X, right positive) components.

*Gage Repeatability and Reproducibility (Gage R&R)*: Parts in a Gage R&R study are the articles being measured; in this study, six parts were included, each of the two leads implanted into the three participants visible in the DRRs. Measurement operations were performed two times by two different trained operators for each lead in each of the three generated DRRs in a randomized order. The same lead electrodes were previously digitized from the 3DCT source data and rostral-caudal and medial-lateral position from vertebral endplate references were assessed and used as a gold standard reference. ANOVA was performed for both Z and X measurements for each electrode (E1 at the most distal lead position through E8 at the most proximal lead position). Part, Operator, and Part*Operator interaction significance threshold was $p<0.05$. Percent Study Variation (%SV) based on six standard deviations was determined for the total Gage R&R and separately for repeatability and reproducibility.

*Accuracy*: Bland-Altman Plots were generated for both Z and X measurements for E1-E8. Mean bias and 95% limits of agreement were plotted and Smallest Detectable Change (SDC) calculated for each electrode.

## Results

A standard intraoperative thoracolumbar radiograph, a representative DRR generated from CT, and an overlay showing points selected between trials and operators are shown in Figure 3. Metal artifact attenuation was successful, but when compared to intraoperative radiographs (Figure 3a) sharpness of electrodes was diminished (Figure 3b). Repeated operator point selections were visibly similar (Figure 3c).

Gage R&R ANOVA outcomes are shown in Table 2. Parts were significantly different for all electrodes. Operator and Part*Operator were not significant for any electrode. Boxplots showing the distribution of %SV for E1-E8 measures are shown in Figure 4; aside from one outlier (X for E8) all %SV accounted for less than 10% of the study variation.

Bland-Altman plots are shown in Figure 5 for Z and X measurements for E1-E8. Mean Z bias increased monotonically from E1 to E8 ranging between -1.8 mm to 0.9 mm with SDC ranging between 3.3 mm (E6) and 5.8 mm (E1). For each individual electrode, individual bias magnitude appeared to increase proportionally to the mean value of measurement. Mean X bias was uniformly negative ranging between -1.0 mm (E5) and -1.6 mm (E7) with SDC ranging between 1.3 mm (E5) and 2.1 mm (E7).

## Discussion

The agreement between operators in assessing lead position was found to be excellent using the radiographic measurement technique with no significant differences between operators nor significant Part*Operator interaction. Furthermore, inclusion of multiple operators and repeated measurements each added a small percentage to the overall study variation (generally less than 10%). The electrodes and vertebral landmarks are clear enough that users can select them consistently.

The SDC was found to be less than the nominal spacing between any two electrodes of the lead (7 mm). The error of the technique is no more than one electrode level for the lead design used in this study. This level of accuracy was also achieved with an imaging modality that is convenient and exposes the patient to a very small radiation dose. If this level of measurement error were to lead a clinician to select an electrode configuration that produces unsatisfactory outcomes, this could potentially be corrected by shifting the target anode or cathode electrode by one position in the array configuration. The SDC found is smaller than what some have reported to be considered significant [11].

The magnitude of bias of the rostral-caudal measurements was found to be directly proportional to the electrode's distance from the reference plane. Basic geometric calculations show that rostral-caudal electrode measurements errors due to out-of-plane tilt of the spine and leads are directly proportional to both the angle of tilt and to the distance of the electrodes from the reference plane (Figure 6). Hofstoetter et al. describe a model incorporating 3D imaging to characterize and correct for this tilt [16]; however such correction would only be relevant if patient position between image modalities were identical. Because the approach calls for selection of a reference plane that is nearest to the centroid of the electrode array, and this should help to limit the magnitude of this error.

This study has several limitations. Patient position during CT scanning was not rigorously controlled to constrain spinal column tilt or rotation, as this data was used retrospectively. However, there are generally no similar tight controls on C-arm orientation when used intraoperatively, so this may have made variance in test parts more realistic. DRR generated from CT data are only models of actual intraoperative radiographs. Key differences between source data and true intraoperative radiographs may affect results. Procedures to implant epidural leads require the patient to lie prone while standard CT imaging is performed with the patient supine. These differences impact relative orientation and position of the lead with respect to the spine and/or the effective film plane. DRRs include the effects of metal artifact;

the artifact is not present in radiographs. This study evaluates measurements performed only by human operators. While repeatability and reproducibility were good in this study, the electrodes and landmarks are likely clear enough that they could be identified with machine learning tools in a more automated approach. Automated point selection will be an enhancement considered in the future.

In conclusion, we have demonstrated that a standardized radiograph technique to quantify lead position performs with acceptable accuracy and the influence of human operators on results is not of concern. With careful control of imaging parameters, measurements of lead position can be made with confidence. Adopting such approaches will yield better information on incidence of lead migration and perhaps adoption of clinical tools and techniques to mitigate or more easily manage it.

**Acknowledgments**

We would like to acknowledge Taylor Trentadue, Ph. D. for providing code for creating digitally reconstructed radiographs. This study was supported by the National Institutes of Health (NINDS Award Number R01NS115877), the Minnesota Office of Higher Education Spinal Cord Injury and Traumatic Brain Injury Research Grant Program (2021 contract: 191542, 2022 contract: 214556) and Mayo Foundation.

**Tables**

Table 1: CT image acquisition protocol parameters

| Parameter | Value |
|---|---|
| Number of participants | 3 |
| CT scanner | Siemens Somatom Definition Edge |
| Gantry rotation time | 1 second |
| Proximal-distal detection coverage (at isocenter) | 57.6 mm |
| X-ray tube voltage | 120 kV |
| Quality reference mAs | 250 |
| Reconstruction kernel | Br64 |
| Voxel size | 0.75 mm slice<br>0.29-0.33 mm in-plane |

Table 2: Gage R&R ANOVA significance for lead electrode measures (Z / X significance) by Part, Operator, Part, and Part*Operator

| Electrode | 1 | 2 | 3 | 4 | 5 | 6 | 7 | 8 |
|---|---|---|---|---|---|---|---|---|
| **Part** | *- - - - P<0.001 for all Z and X - - - -* | | | | | | | |
| **Operator** | 0.79/ 0.92 | 0.67/ 0.66 | 0.88/ 0.90 | 0.94/ 0.37 | 0.90/ 0.21 | 0.91/ 0.15 | 0.48/ 0.32 | 0.30/ 0.22 |
| **Part* Operator** | 0.47/ 0.93 | 0.57/ 0.25 | 0.39/ 0.89 | 0.84/ 0.50 | 0.84/ 0.35 | 0.68/ 0.62 | 0.88/ 0.26 | 0.95/ 0.67 |

## Figures

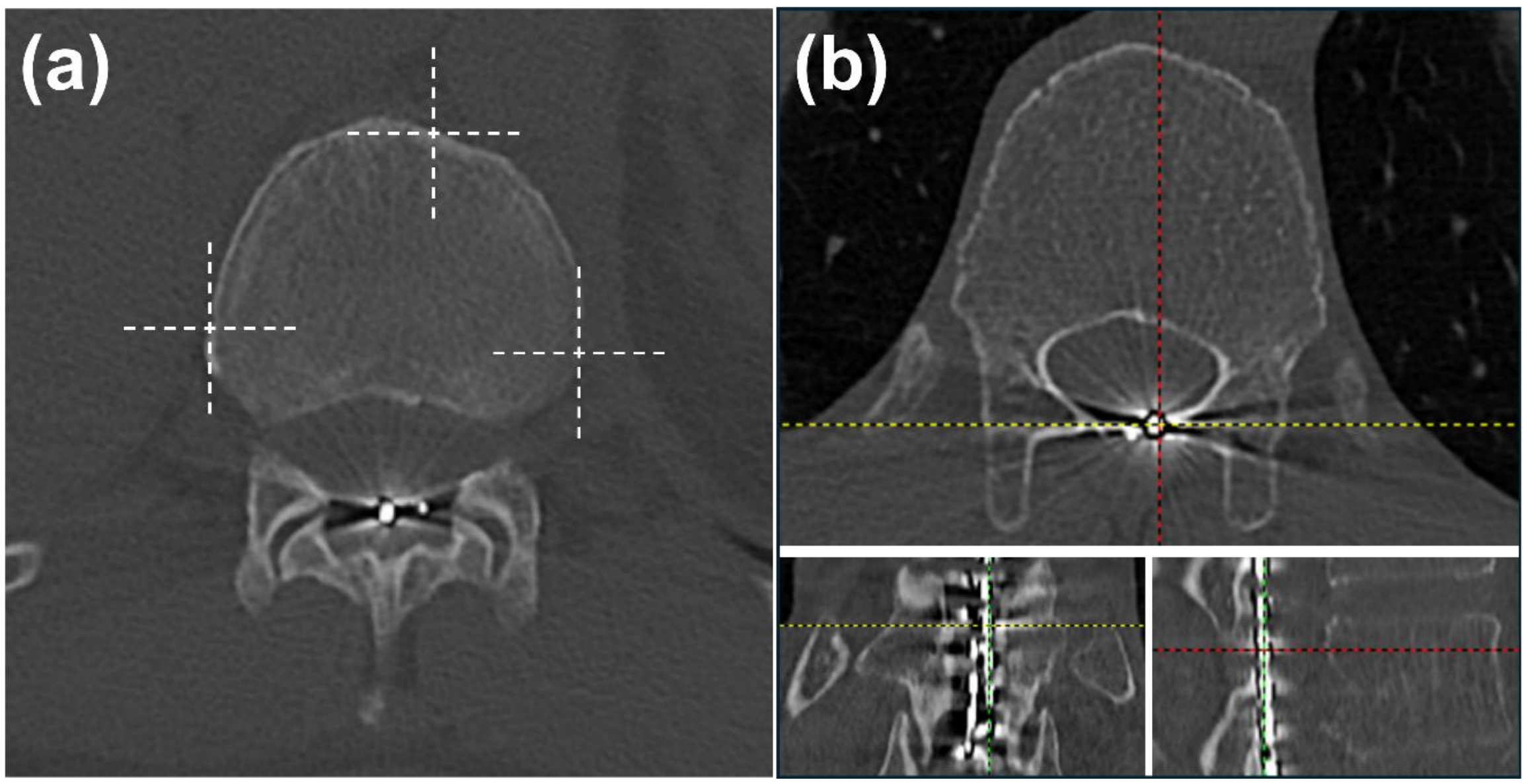

Figure 1: Gold standard measurements of lead position were defined from patient post-operative CT by digitizing a) three points on the vertebral superior endplates to define reference planes and b) the centroid of each lead electrode.

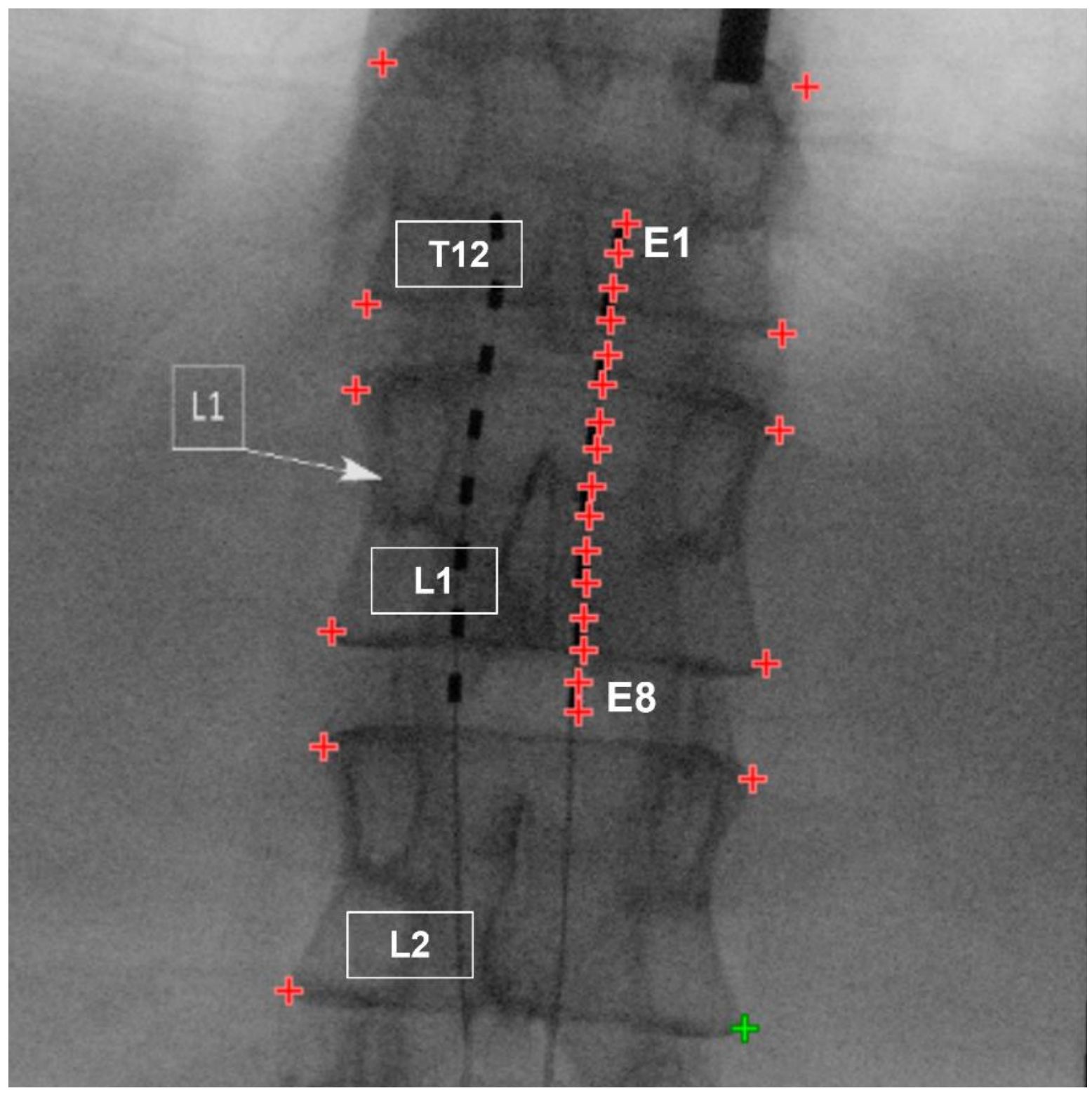

Figure 2: The radiographic measurement protocol required digitizing points defining the superior and inferior aspect of each lead electrode and the four corners bounding the visible region of the anterior vertebral bodies for T12, L1 and L2.

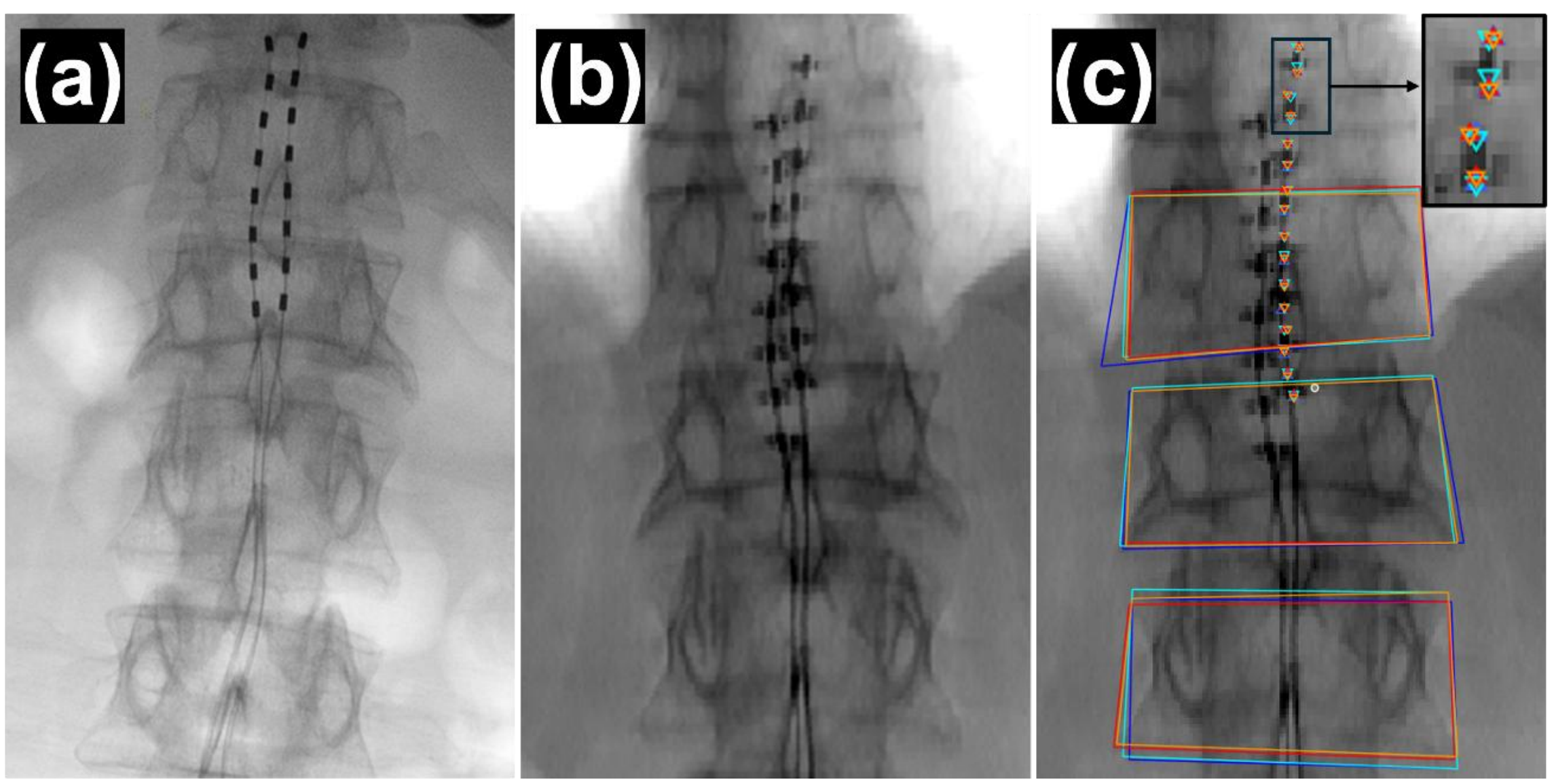

Figure 3: Representative examples of a) actual intraoperative plane radiograph of lead and surrounding anatomy obtained with a Phillips Medical Systems Veradius Unity PH Mobile C-Arm, b) a digitally reconstructed radiograph created from patient CT with metal artifact attenuation, and c) a comparison between repetitive and interoperator points selected (blue/cyan: operator 1, red/orange: operator 2) on a DRR as input to radiographic measurements.

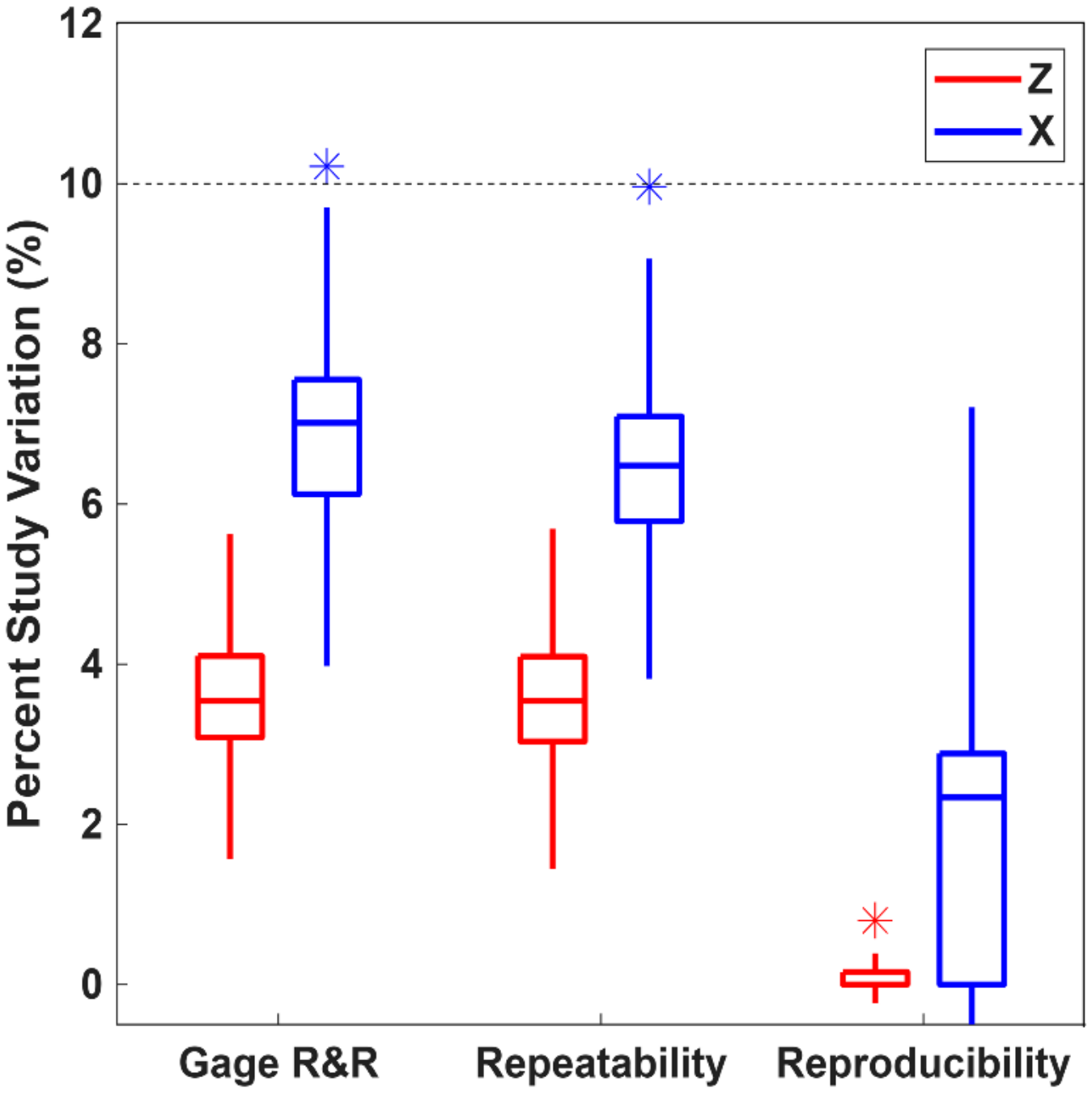

Figure 4: Boxplot showing distribution of Gage Repeatability and Reproducibility, Repeatability, and Reproducibility contributions to study variability across Z (red) and X (blue) lead electrode measurements. Percent Study Variation for each of these was generally less than 10% of the total study variation.

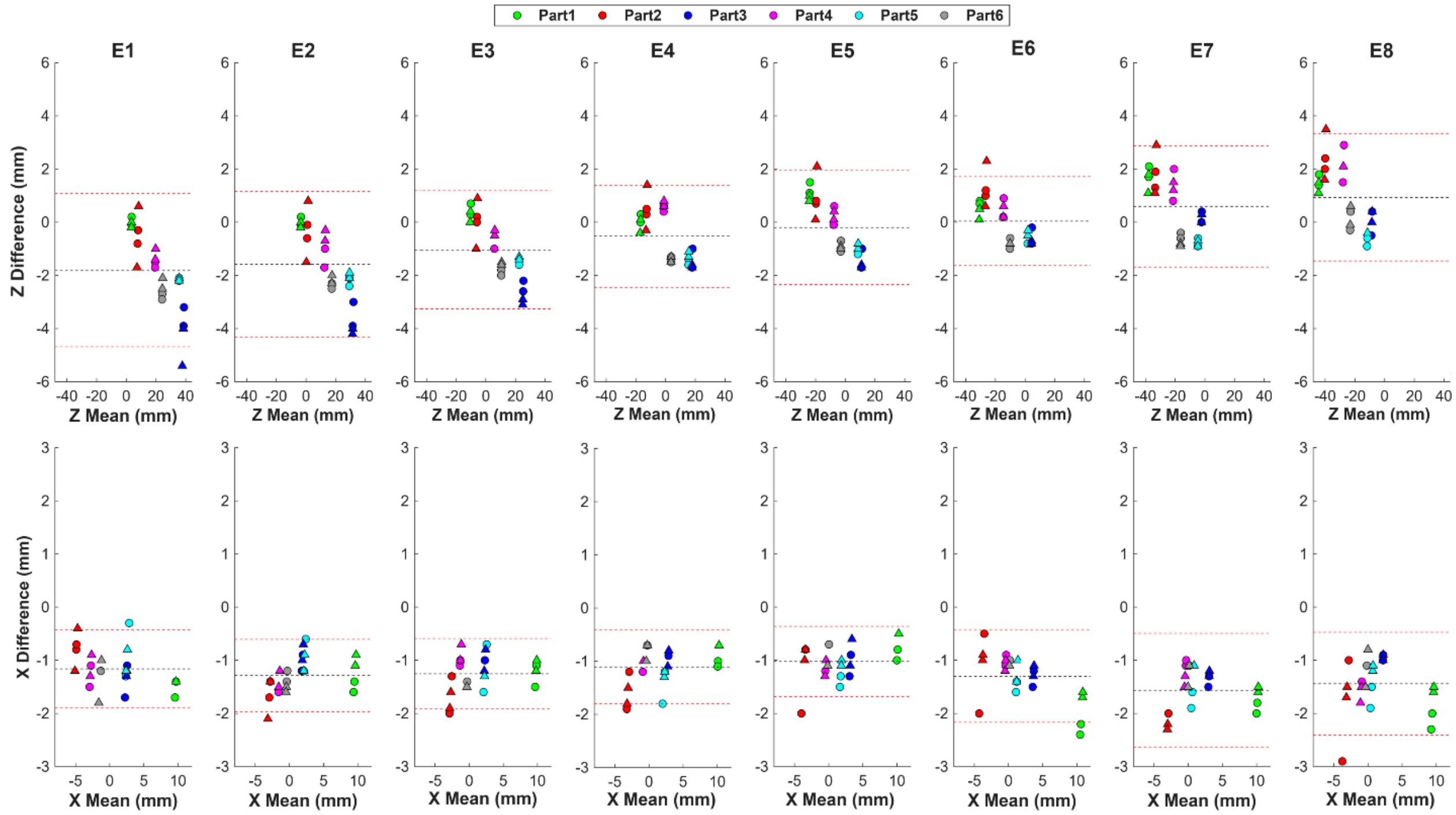

Figure 5: Bland-Altman plots for measurements of Z (top row) and X (bottom row) electrode positions for E1-E8. Mean biases are indicated with black lines, and 95% Limits of Agreement are shown with red lines.

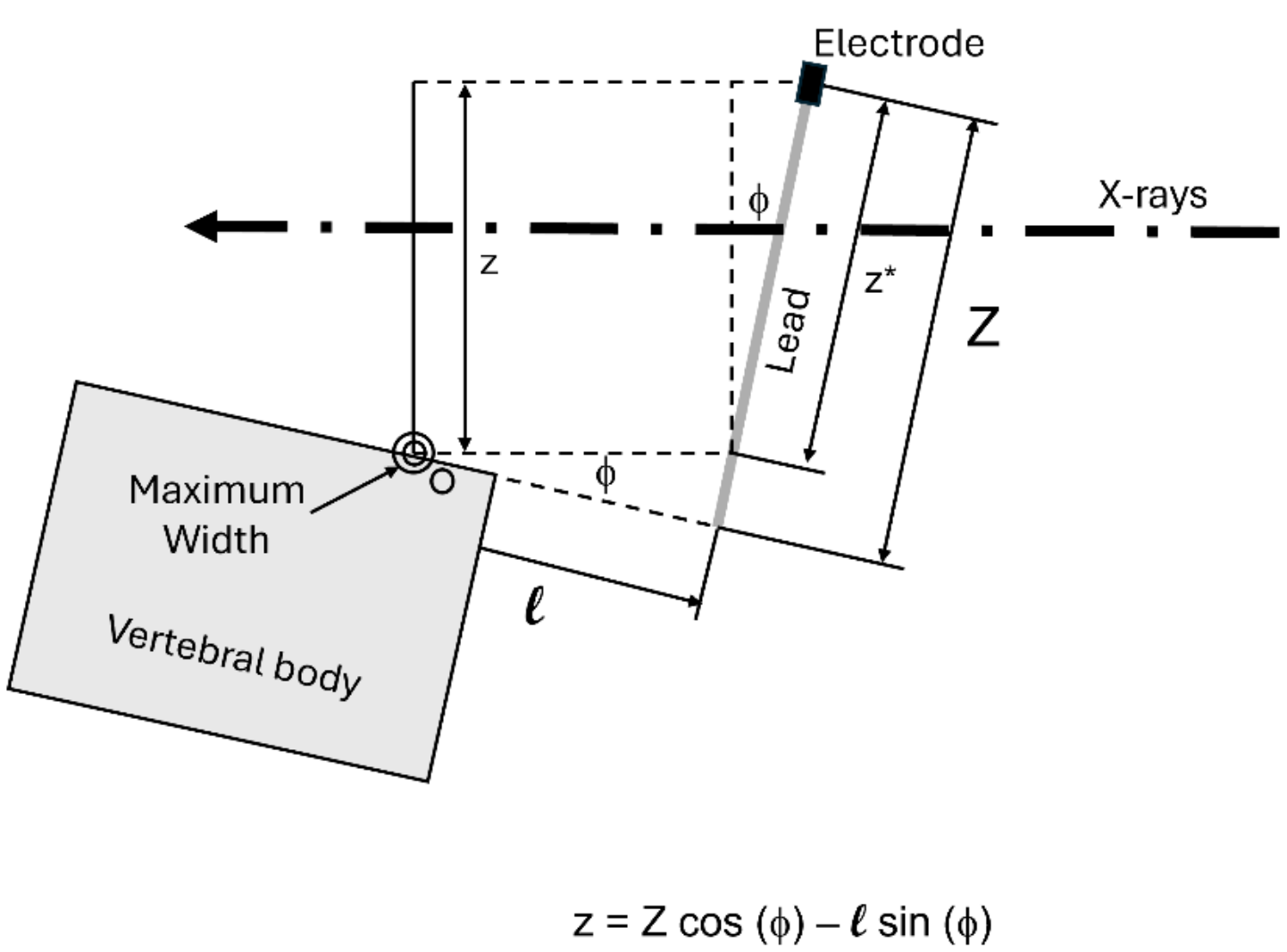

$$z = Z\cos(\phi) - \ell\sin(\phi)$$

Figure 6: Effects of out-of-plane tilt can be assessed from basic geometric relationships considering some angle of tilt, f, the lead is some distance, **l**, from the widest point of the vertebral body project in the radiograph, Z is the true distance from the vertebral reference plane and z is the observed measurement.